\documentclass[preprint2]{aastex}

\begin{document}

\title{Physics of a partially ionized gas relevant to  galaxy
  formation simulations -- the ionization potential energy reservoir}

\author{B.~Vandenbroucke, S.~De~Rijcke, J.~Schroyen}  
\affil{Dept. Physics \& Astronomy, Ghent University, Krijgslaan 281, S9, 9000 Gent, Belgium}
\and
\author{N.~Jachowicz}
\affil{Dept. Physics \& Astronomy, Ghent University, Proeftuinstraat 86, 9000 Gent, Belgium}

\begin{abstract}
Simulation codes for galaxy formation and evolution take on board as
many physical processes as possible beyond the standard gravitational
and hydrodynamical physics. Most of this extra physics takes place
below the resolution level of the simulations and is added in a
``sub-grid'' fashion. However, these sub-grid processes affect the
macroscopic hydrodynamical properties of the gas and thus couple to
the ``on-grid'' physics that is explicitly integrated during the
simulation.

In this paper, we focus on the link between partial ionization and
the hydrodynamical equations. We show that the energy stored in ions
and free electrons constitutes a potential energy term which breaks
the linear dependence of the internal energy on temperature. Correctly
taking into account ionization hence requires modifying both the
equation of state and the energy-temperature relation. We
implemented these changes in the cosmological simulation code
\textsc{Gadget2}.

As an example of the effects of these changes, we study the
propagation of Sedov-Taylor shock waves through an ionizing
medium. This serves as a proxy for the absorption of supernova
feedback energy by the interstellar medium. Depending on the density
and temperature of the surrounding gas, we find that up to 50\% of the
feedback energy is spent ionizing the gas rather than heating it.
Thus, it can be expected that properly taking into account ionization effects
in galaxy evolution simulations will drastically reduce the effects of
thermal feedback. To the best of our knowledge, this potential energy term
is not used in current simulations of galaxy formation and evolution.
\end{abstract}


\maketitle

\section{Introduction}
Even with the computing power currently available, it is still
impossible to include {\em all} relevant physical processes in
numerical simulations of galaxy formation and to explicitly solve {\em
  all} the relevant equations. Therefore, recourse is often taken to
splitting the physics into a numerically manageable `macroscopic'
part, whose equations are integrated explicitly during a simulation,
and a numerically intractable `microscopic' part, whose equations are
not solved explicitly \citep{springel_gadget2, ramses, arepo}. Out of
necessity, these microscopic extensions usually offer a
phenomenological or heuristic description of much more complex
physical processes. Examples are complex nuclear reaction networks
\citep{stellar_gadget} and radiative cooling and heating of a
multi-phase, multi-component gas \citep{shen,anninos}. Clearly, it is
mandatory for the assumptions underlying the microscopic physics to be
compatible with those on which the macroscopic physics is based.  A
mismatch between the microscopic and macroscopic levels could hamper the
reliability of numerical hydrodynamical simulations in a priori
unknown ways.

In this paper, we focus on the connection between the microscopic and
macroscopic physics in hydrodynamical simulations of an ionizing
gas. Our main interest is the study of the low-density interstellar
medium of galaxies, subject to heating by supernova explosions and
the cosmic ultraviolet background. The lowest-order gas model is the
so-called ideal gas. Extending on this, the use of more general
equations of state is widespread in the literature. In
\cite{Choi_Wiita} a temperature-dependent treatment of the adiabatic
index of a simple proton-electron plasma was proposed, while in
\cite{Marek} a study of a mixture of baryonic gas and neutrinos was
performed. Likewise, the energy equation has been adapted to different
physical circumstances. For instance, a gas is heated when irradiated
by ionizing UV radiation while recombination is a source of cooling
due to the radiative loss of the kinetic energy of each captured
electron \citep{Mizuta}.

However, because of the energy spent freeing an electron, each
ion-electron pair constitutes a small amount of potential
energy. Therefore, a partially ionized gas has an internal potential
energy reservoir at its disposal that can absorb and release
energy. In a gas composed of only Hydrogen and Helium and sufficiently dense
to be in thermodynamic equilibrium, this potential energy
reservoir can straightforwardly be taken into account using Saha's
equation to determine the ionization equilibrium and the
Stefan-Boltzmann law to treat both cooling and heating
\citep{stamatellos,mihalasx2}. This approximation is valid for
circumstellar discs and star-forming molecular cores but it does not
apply to a multi-component, optically thin gas such as the
interstellar medium of galaxies. This requires a more rigorous
numerical calculation of the ionization equilibrium and the
corresponding potential energy stored in ions while also using a
realistic treatment of gas cooling and heating. 

The kinetic energy of the free electrons and the heating of gas,
e.g.~by ionizing radiation, have already been included in many
state-of-the-art hydrodynamics codes, but this potential energy
reservoir was to the best of our knowledge never before taken
into account in cosmological or galaxy evolution simulations. In this
paper, we show how this potential energy reservoir can be added to
the microscopic physics of a numerical simulation and we investigate
its effects.

\section{Modified hydrodynamics}

The internal energy of a partially ionized multi-component gas is
given by
\begin{eqnarray}
\rho u &=& \frac{3}{2} \left(\sum_X n_X + n_e(T) \right) kT \nonumber
\\ &&+ \sum_X \left[\sum_{i} n_{X,i}(T) \left( \sum_{j=1}^{i}
  \varepsilon_{X,j} \right) \right]. \label{rhou}
\end{eqnarray}
Here, $n_{X,i}(T)$ is the number density of the $i^\textrm{th}$ ion of
element $X$, $n_e$ is the free electron density, and
$\varepsilon_{X,i}$ is the energy required to produce this ion by
removing an electron from the $(i-1)^{\textrm{th}}$ ion, ground state
to ground state. From charge conservation it follows that $n_e(T) =
\sum_{i} i \,n_{X,i}(T)$. The second term in Eq. (\ref{rhou}) gives
the potential energy reservoir associated with the ionization. The
temperature-dependent density of free electrons also affects the
equation of state.

The system of hydrodynamical equations is closed by introducing an
equation of state, which in this case reads
\begin{equation}
p = \left( \sum_X n_X + n_e(T) \right) kT = \frac{\rho k T}{\mu(T)},
\end{equation}
and which is easily implemented by using a temperature-dependent mean
particle mass $\mu(T)$. It is clear from these equations that the
relation $p = (\gamma - 1)\rho u$, which is generally used to simplify
the hydrodynamical equations \citep{springel_entropy} no longer
holds. This means that (a) the entropy based formulation of
hydrodynamics is no longer possible and, (b) temperature becomes a
hydrodynamical variable in the system and has to be integrated along
with the equations of motion.

\subsection{Energy-temperature dependence and the adiabatic index}

We have numerically calculated and compiled the ionization
equilibrium, and the cooling and heating rates of gases with a wide
range of compositions, densities, temperatures, and heating sources
\citep{de13}.  For this work, we have extended the capabilities of
\textsc{ChiantiPy}, a Python interface to the \textsc{Chianti} atomic
database \citep{chianti}. For all ions, we use the recombination
rates, collisional ionization rates, and energy level populations
provided by \textsc{ChiantiPy}. Photo-ionization cross-sections are
adopted from \cite{ve96} and integrated over the adopted stellar and
cosmic UV- backgrounds in order to obtain the photo-heating and
radiative cooling rates. The latter are required for the integration
of the energy equation. With the ionization equilibrium in hand, we
can evaluate Eq.~(\ref{rhou}) and infer the gas temperature from the
internal energy. The temperature derivatives of Eq.~(\ref{rhou}) at
constant density and at constant volume then yield the temperature
dependent adiabatic index $\gamma(T)$ of the gas.

\begin{figure}[t]
\includegraphics[width=0.45\textwidth]{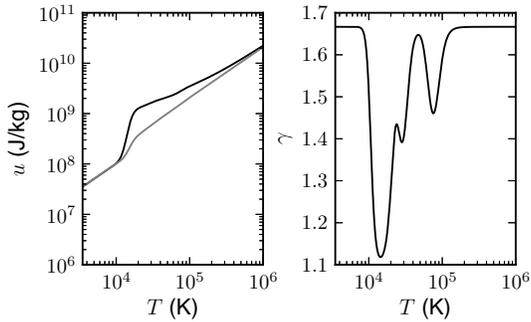}
\caption{\label{fig1}Specific internal energy (left) and
  adiabatic index $\gamma$ (right) as a function of temperature
  for a gas consisting of 92\% H and 8\% He. The gray line on the left
  plot traces the internal energy without taking into account the
  potential energy reservoir.}
\end{figure}

The adiabatic index and internal energy are plotted in Fig. \ref{fig1}
for of a gas consisting of 92\% H and 8\% He (by particle number),
mimicking the unenriched interstellar medium. The transitions between
the different ionization stages of H and He are clearly visible as
steep drops of the adiabatic index. For comparison, we also plotted
the internal energy without the potential energy reservoir (gray
curve). Taking into account the potential energy reservoir associated
with the ionization significantly alters the hydrodynamical properties
of the gas; around 15,000~K, where $\gamma \sim 1.1$, the gas behaves
almost isothermally:~the ionizing hydrogen provides the gas with a
large number of additional degrees of freedom. Heavier elements in the
interstellar medium usually have abundances that are significantly
smaller than those of H and He, and have a negligible impact on the
adiabatic index. This was verified by calculating the adiabatic index
with \textsc{Chianti}, both with and without metals present. Even for
solar abundance ratios \citep{grevesse}, both calculations yield
almost identical results.

\subsection{Excited states}
So far, we assumed all ions to be in their ground states. For dense
gases, this approximation can be invalid. It is straightforward to
extend expression (\ref{rhou}) by including a summation over each
ion's energy level occupations. This more general
treatment can then be used to check if the approximation holds.  For a given ion,
excited states usually only become populated at temperatures where
further ionization also gains importance and the ion's density
decreases steeply. Hence, their effect is completely dominated by the
ionization effect. However, for elements like O and Fe having
low-lying excited states, the picture is less straightforward. We
calculated the adiabatic index, with and without taking into account
excited states, and found essentially identical results, even for
solar abundance ratios. Only for a pure O or Fe gas does the
difference amount to a few percent at most. This justifies the use of
this approximation.

\section{Sedov-Taylor blast wave}

The effect of our advanced treatment of ionization
on the gas dynamics can be illustrated by means of a classical test
case for hydrodynamical simulation codes:~the Sedov-Taylor blast
wave. The problem consists of a homogeneous gas cloud, in which a
spherical shock wave propagates radially outward. For a gas with
negligible internal energy and constant adiabatic index, the radius of
the shock wave as a function of time is given by
\begin{equation}
r_s = \left( \frac{E_0}{\alpha(\gamma) \rho_1} \right)^{1/5}
t^{2/5},
\end{equation}
where $\alpha(\gamma)$ is a numerical factor which can be calculated
by imposing energy conservation, $E_0$ is the total energy of the
shock, and $\rho_1$ is the density of the ambient gas
\citep{sedov}. $\alpha$ is a monotonically decreasing function of
$\gamma$, hence the shock wave propagates more slowly in a medium with
smaller $\gamma$ or, equivalently, with more internal degrees of
freedom per particle. If the gas has a non-negligible internal energy,
the counter-pressure exerted by the ambient gas on the shock front
must be taken into account. The analytical solution is then somewhat
more involved but leads to the same conclusion:~ionization is
anticipated to slow down the propagation of expanding shock
waves.

\begin{figure*}
\includegraphics[width=0.95\textwidth]{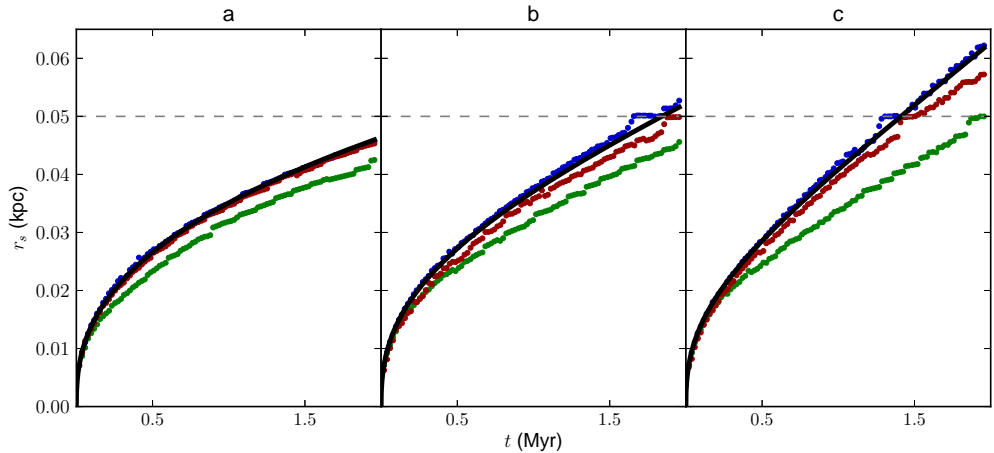}
\caption{\label{fig_shock_profile}Shock wave radius as a function of
  time for a gas with a density of $\rho = 81$~amu~cm$^{-3}$ and
    an initial temperature (a) 100~K, (b) 5,000~K, and (c)
  15,000~K. The gray dots (blue dots in the online version) represent
  the simulation values with the old gas physics, the black dots
  (green dots in the online version) are the values for the new gas
  physics, the full line is the analytical solution following
  \protect\cite{sedov}. The gray crosses (red dots in the online
  version) represent simulation values obtained by omitting the
  potential energy term, but with taking into account the change in
  particle number caused by the ionization.}
\end{figure*}

\begin{figure*}
\includegraphics[width=0.95\textwidth]{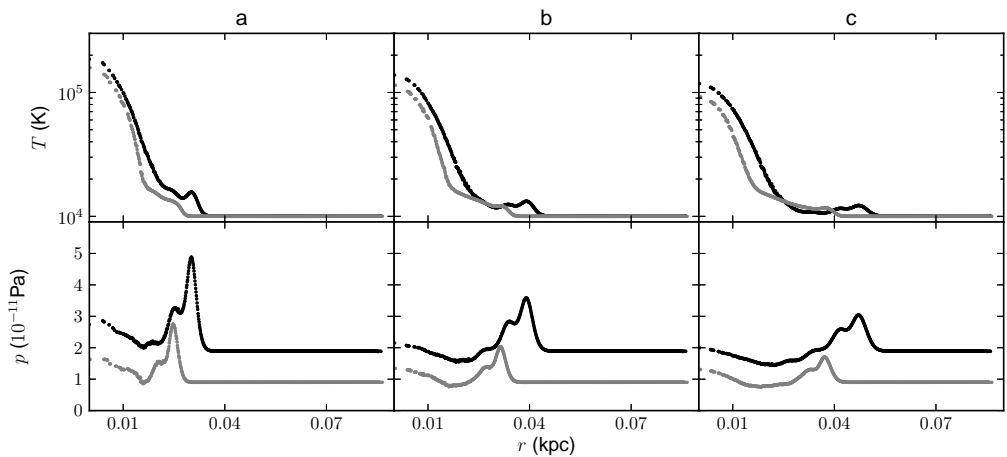}
\caption{\label{fig_temp}Temperature and pressure profile of the shock
  wave traveling through a gas with density $\rho =
    81$~amu~cm$^{-3}$ and an initial temperature of 10,000~K for
  timesteps (a) 0.6~Myr, (b) 1.0~Myr, and (c) 1.4~Myr. The black dots
  represent the old gas physics, the gray ones the new gas
  physics.}
\end{figure*}

\subsection{Simulations}
In order to obtain a full understanding of the influence of ionization
on the propagation of a shock wave in numerical computations, we
conducted a series of SPH-simulations, using a modified version of the
cosmological simulation code \textsc{Gadget2}
\citep{springel_gadget2,valcke}. We employ a prescription for artificial
thermal conductivity \citep{price}, that is used to smooth the
initial central energy peak during the first few timesteps, and a
timestep criterion along the lines of that proposed by \cite{arepo},
which appropriately increases the number of timesteps used to
integrate the motions for a particle just ahead of the shock wave
\citep{saitoh_makino}. We note that these modifications are crucial to
perform high resolution simulations of a Sedov-Taylor blast wave
without severe energy losses during the first few timesteps, and
without a potentially fatal mistreatment of the initial shock. We ran
simulations with two different types of the code:~one version which
treated the gas as an ideal, fully ionized H plasma with a constant
adiabatic index (hereafter denoted as ``old'' gas physics) and one
version where the correct energy and pressure relations are used
(dubbed ``new'' gas physics). In order to implement the new gas
physics, we had to replace the entropy equation in \textsc{Gadget2}
with the energy equation, as already noted above.

As initial conditions for our simulations, we used a cubic box with
periodic boundary conditions and side 0.1~kpc, wherein a total gas
mass of $2\times 10^{6}$ solar masses was sampled using $300,000$
randomly distributed SPH-particles. For 2~Gyr, these particles were
first evolved using \textsc{Gadget2} in order to allow the system to
reach an equilibrium density of $\rho = 81$~amu~cm$^{-3}$. This is a
value typical for a cold, dense star-forming cloud in a galaxy
evolution simulation (see for example
\cite{governato,cloetosselaer,schroyen2013}, who use a density
threshold for star formation of up to
100~amu~cm$^{-3}$). Various initial temperatures, ranging from 100~K
to 15,000~K, were used, in order to demonstrate the effect of H
ionization. After the system had reached an equilibrium, an energy of
$10^{52}$~erg was injected, mimicking the simultaneous energy
injection of about 10 supernovae (the typical stellar feedback used in
for example \cite{schroyen,cloetosselaer}), by giving this energy to
the particle closest to the box center. The system was then evolved
using \textsc{Gadget2}.

In order to assess the combined effects of initial temperature
  and density on our results, we also simulated the expansion of shock
  waves through gas with densities of 0.81~amu~cm$^{-3}$ and
  0.0081~amu~cm$^{-3}$, and for the same initial temperatures as listed
  above. These simulations were set up by sampling a total gas mass
which was respectively a factor 100 and a factor 10,000 smaller, using
the same procedure as above, within an equally large box. This
should give us an idea of how the new gas physics scales with density
and together with the different temperatures explored should cover a
broad range of scenarios occurring in galaxy evolution
simulations.

\section{Results}
To characterize the resulting shock wave we analysed its radial
distance from the box center, which is well quantified by the radius
of the SPH particle with the highest density, as a function of
time. The results are shown in Fig.~\ref{fig_shock_profile}, for both
the old (gray dots, blue in the online version) and the new gas physics (black dots, green in the online version), and are compared with the analytical
solution with a constant adiabatic index $\gamma=\frac{5}{3}$ (full black line). The latter is seen to agree very well with the old gas
physics, except for the small deviation at the end of the simulation
that is caused by boundary effects as the shock wave hits the periodic
boundary of the box. There is a clear difference between simulations
using the old and the new gas physics; in the latter, the shock wave
travels substantially slower. This is due to the fact that part of the
shock wave energy is converted into (potential) ionization energy.

The effect is also partially due to the change in mean particle mass,
as this will cause the pressure to be lower at temperatures below
10,000~K. To illustrate the contributions of both the change in mean
particle mass and the potential energy reservoir, we also plotted the
results for simulations only implementing the former in
Fig.~\ref{fig_shock_profile} (gray crosses, red dots in the online
version).

To further illustrate the damping of the shock wave, we plotted temperature and
pressure profiles for both versions of the gas physics in
Fig.~\ref{fig_temp}, for a simulation with initial temperature
$T=10,000$~K. The shock wave is visible as the outward propagating
temperature and density peak. For the old gas physics (black curves)
this peak clearly stands out while for the new gas physics (gray curves) it
is strongly suppressed. In the latter case, a substantially smaller
fraction of the shock wave's energy goes into heating the gas and is
instead spent ionizing the gas. As a result, the pressure along the
shock front is lower for the new gas physics, which results in a
slower shock wave propagation. Note also that the overall pressure is
lower for the new gas physics, as a result of the higher mean particle
mass $\mu(T)$ at the initial temperature.  

\begin{figure}[!t]
\includegraphics[width=0.5\textwidth]{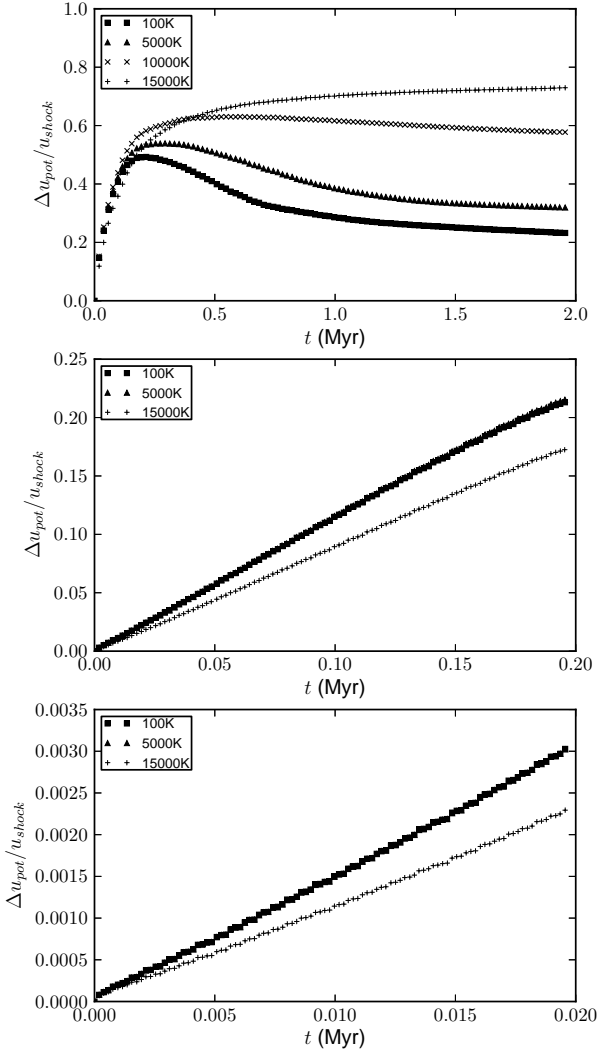}
\caption{\label{fig_ratio}The ratio of the total potential energy
  change and the injected energy as a function of time for different temperature values close to the ionization temperature of H. Top: gas density of 81~amu~cm$^{-3}$, middle: gas density of 0.81~amu~cm$^{-3}$, bottom: gas density of 0.0081~amu~cm$^{-3}$.}
\end{figure}

We plot the ratio of the increase in potential energy to the initial
shock wave energy in Fig. \ref{fig_ratio}. For the high density
simulations, at least 50 \% of the initially injected energy is
converted into potential energy for all temperatures. As the gas
expands, cools, and recombines after the passage of the shock wave,
the potential energy can decline again. For lower densities, this
effect is largely compensated by the large increase in shockwave
temperature resulting from the much lower density. At
0.81~amu~cm$^{-3}$, up to 20 \% of the shockwave energy is converted
into potential energy by the time the shock reaches the box
  edge. The effect is strongest for low temperatures, as in this case
a larger part of the potential energy reservoir is still available for
excitation. At the lowest density of 0.0081~amu~cm$^{-3}$, the total
shockwave energy is orders of magnitude higher than the combined
contribution of all potential energy terms, so the ratio stays close
to zero for the part of the evolution we consider. At these lower
  densities, the potential energy is roughly proportional to the
  amount of gas that has been ionized and can hence be expected to be
  proportional to the volume within the spherically expanding shock
  wave. Since $r_s \propto t^{2/5}$ it follows that we expect $\Delta
  u_{pot}/u_{shock} \propto t^{6/5}$. This quasi-linear behaviour is
  indeed observed in the middle and bottom panels of
  Fig.~\ref{fig_ratio}.

\section{Discussion and conclusions}
The damping of shock waves has important repercussions in our specific
case of galaxy evolution simulations, where supernova feedback is
often injected into the interstellar medium as thermal energy. The
Sedov-Taylor experiments described here give a fair idea of the
effects that can be expected in full-fledged galaxy
simulations. Including the new gas physics described here, the
expanding bubbles of hot, dilute gas that surround stellar nurseries
will propagate more slowly into the surrounding interstellar medium.
Moreover, the interstellar gas will be heated less by the supernovae,
as part of the feedback energy will be spent on ionizing the gas. A
detailed account of the precise effect of our improved description of
gas dynamics on simulations of galaxy evolution is the subject of
future work (B. Vandenbroucke et al. 2013, in preparation).

To conclude, we have shown how the effects of ionization on the
`macroscopic' dynamics of a low-density gas can be taken into account
numerically using the proper `microscopic' physics. We have used
hydrodynamical simulations to illustrate that employing the improved
equation of state and thermal energy density has a significant effect
on the propagation of spherically expanding shock waves in a Hydrogen
gas. The ionization opens up new degrees of freedom that make the gas
behave essentially isothermally when Hydrogen ionizes or
recombines. This significantly damps shock waves. We have shown that
this limits the efficiency with which supernova-blown shock waves
inject energy into the interstellar medium to less than 50 \%. We therefore strongly encourage the use of the potential energy term
in the expression for the thermal energy density, not only in the case
of ionization, but also for other astro-chemical reactions.

We have made tables available online containing the mean particle mass and the specific energy per unit mass as a function of temperature\footnote{http://users.ugent.be/$\sim$sdrijcke}. These are the same tables which were used to perform the simulations in this paper. More details about these tables and the cooling tables associated with them can be found in \cite{de13}.

\acknowledgments

We acknowledge \textsc{Chianti}, a collaborative project involving
researchers at NRL (USA), RAL (UK), and the Universities of: Cambridge
(UK), George Mason (USA), and Florence (Italy).

\end{document}